\documentclass[conference]{IEEEtran}
\IEEEoverridecommandlockouts
\usepackage{cite}
\usepackage{amsmath,amssymb,amsfonts}
\usepackage{subcaption}
\usepackage{graphicx}
\usepackage{textcomp}
\usepackage{xcolor}
\usepackage{booktabs}
\usepackage{hyperref}
\usepackage{url}

\begin{document}

\title{Benchmarking Storage Systems for Machine Learning Workloads Using NIO Bench}

\author{
\IEEEauthorblockN{
Jonathan W. Morris, Ionut Mistreanu, Connor Louie
}
\IEEEauthorblockA{
\textit{University of California, Santa Cruz} \\
\{jowemorr, imistrea, cdlouie\}@ucsc.edu
}
}
\maketitle

\begin{abstract}
Machine learning training workloads place unique demands on storage systems, yet most existing benchmarks focus on computational throughput rather than file system I/O behavior. We present a benchmarking framework, Neural I/O Benchmark (NIO Bench), that characterizes storage access patterns across six diverse ML model architectures: Language Transformers, Vision Transformers, Diffusion Models, Spiking Neural Networks, Artificial Neural Networks, and Reinforcement Learning. Our framework employs a two-layer tracing approach combining Python-level I/O hooks for semantic phase context with Linux \texttt{strace} for complete syscall coverage including DataLoader worker subprocesses. We evaluate all six models on a Nautilus Kubernetes cluster with Ceph distributed file system. Our results reveal that I/O is heavily concentrated in data preparation, model loading, and model checkpointing. We also found that training is compute-bound rather than data-bound once data is staged, and that storage access follows an extreme power law where fewer than 10\% of files account for over 90\% of bytes transferred, and that read tail latency from cache misses on distributed storage is the primary storage bottleneck. These findings suggest that storage systems optimized for ML should prioritize aggressive data prefetching, page cache pinning, and efficient handling of bursty checkpoint writes.
\end{abstract}


\section*{Artifact Availability}
The source code for NIO Bench, including the tracing framework, model training scripts, and analysis scripts used to generate the figures in this paper, is available at:
\url{https://github.com/JonathanWMorris/storage-system-ML-benchmark}.

\section{Introduction}

Storage systems are an increasingly significant bottleneck in modern computing, particularly for Machine Learning (ML) workloads. As ML models grow in scale, from millions to billions of parameters, the demands placed on storage systems during training have intensified. \cite{bahri_explaining_2024} These demands include reading large datasets, loading pretrained model weights, writing periodic checkpoints, and exporting trained models. Unlike traditional workloads such as databases or web servers, ML training exhibits distinct I/O patterns: bursty writes during check-pointing, sustained sequential reads during data loading, and long idle periods during GPU-bound computation.

Despite the growing importance of storage performance for ML, current file systems are not optimized for these access patterns. Most are designed for general-purpose workloads with assumptions about access frequency, file sizes, and read/write ratios that do not hold for ML training. Understanding how different model architectures interact with storage is a prerequisite for designing better storage systems for ML.

The goal of this work is as follows: (1) Create a comprehensive I/O benchmark, named Neural I/O Benchmark (NIO Bench), for popular machine learning workflows. (2) Characterize I/O patterns across different ML model architectures on a Kubernetes cluster. (3) Identify common patterns and model-specific behaviors. (4) Provide insights for potential storage system modifications to improve ML storage systems based on our results.

\section{Related Work}

Several benchmarking suites exist for evaluating ML system performance. Fathom~\cite{adolfFathomReferenceWorkloads2016} provides a set of reference workloads spanning different deep learning architectures. BenchNN~\cite{shiBenchmarkingStateoftheArtDeep2017} benchmarks neural network training performance across hardware platforms. NVIDIA MLPerf~\cite{reddiMLPerfInferenceBenchmark2020} is the industry standard for measuring ML training and inference throughput. However, all three focus primarily on computational performance, such as GPU utilization, throughput, and time-to-accuracy, and do not characterize file system I/O behavior.

The closest related work is DLIO (Deep Learning I/O)~\cite{devarajanDLIODataCentricBenchmark2021}, which specifically targets I/O characterization for deep learning. DLIO profiles data loading pipelines and analyzes I/O bottlenecks during training. However, DLIO focuses on a limited set of model architectures---primarily CNNs \cite{oshea_introduction_2015}, RNNs \cite{schmidt_recurrent_2019}, and basic ANNs \cite{rosenblatt_perceptron_1958}---that represent an older generation of ML models. Modern architectures such as vision transformers, large language models, diffusion models, and spiking neural networks have substantially different I/O profiles and are not covered by existing benchmarks.

Our work extends the DLIO approach to a broader set of contemporary model architectures and introduces a two-layer tracing framework that captures both application-level semantics and complete syscall-level I/O, which we cover in section \ref{Two-Layer Tracing Architecture}.

\section{Benchmark Design}
\label{Benchmark Design}

\subsection{Model Selection}
\label{Model Selection}

We selected six model architectures that span the breadth of modern ML workloads, each representing a distinct category of I/O behavior:

\begin{enumerate}
\item \textbf{Language Transformer (LT):} Fine-tuning Llama-3.1-8B \cite{grattafiori_llama_2024} with LoRA/QLoRA \cite{hu_lora_2021} on Amazon Reviews 2023 \cite{hou_bridging_2024}. This model has the most complex I/O pipeline with five distinct stages: data staging, tokenization, sharding, training with periodic checkpointing, and merged model export. It produces the most diverse I/O patterns of all models tested. For our tests, we only finetuned on a small sample of the Amazon Reviews dataset to keep the benchmarking time for NIO Bench manageable.

\item \textbf{Artificial Neural Network (ANN):} A three-layer MLP with ReLU \cite{agarap_deep_2019} activations trained on the UCI Adult Income dataset ($\sim$3~MB). This serves as our low-I/O baseline, tabular data that fits entirely in memory after initial download.

\item \textbf{Spiking Neural Network (SNN):} A two-layer spiking MLP using Leaky Integrate-and-Fire neurons \cite{lu_linear_2022}(via SNNTorch\cite{eshraghian_training_2023}) trained on N-MNIST ($\sim$1~GB). We chose the N-MNIST dataset to model IOT Computer Vision applications of SNNs. Temporal processing over 20 time bins increases compute per sample while maintaining standard data loading patterns. 

\item \textbf{Vision Transformer (ViT):} ViT-Base/16 \cite{dosovitskiy_image_2021} (via timm) fine-tuned for multi-label classification on COCO 2017 \cite{lin_microsoft_2015} ($\sim$20~GB, 118K images). This model is I/O-intensive during training, producing millions of small JPEG reads through DataLoader worker subprocesses each epoch.

\item \textbf{Diffusion Model:} Fine-tuning Stable Diffusion \cite{rombach_high-resolution_2022} v1.5's UNet2D \cite{ronneberger_u-net_2015} ($\sim$860M parameters) on COCO captions using the DDPM \cite{ho_denoising_2020} noise prediction objective. Heavy initial I/O from loading $\sim$4~GB of pretrained weights plus the 20~GB dataset, with large periodic checkpoint writes during training.

\item \textbf{Reinforcement Learning (RL):} PPO\cite{schulman_proximal_2017} with CNN policy (via Stable Baselines3) on Atari Learning Environment's Space Invaders \cite{mnih_playing_2013} with 4 parallel environments. Training data is generated at runtime through agent-environment interaction, making this our lowest-I/O workload with no dataset reads during training.
\end{enumerate}

The models were chosen based on popularity of recent ML workloads. Although ANNs are not as data intensive as LLMs or ViTs, we still included them for completeness of NIO Bench. SNNs were included in NIO Bench as it is an emergent technology with promising applications for low power and specialized neuromorphic devices, and because we wanted to see if backpropagation through time used by SNNs have different I/O characteristics to traditional backpropagation used by ANNs.
We chose the COCO dataset for Diffusion Model and ViT rather than larger models like ImageNet \cite{russakovsky_imagenet_2015} because COCO does not require permission to download and use their dataset. In addition, we also preferred COCO as it contains more images than CIFAR \cite{noauthor_cifar-10_nodate}dataset, making it the ideal vision dataset for NIO Bench.

\subsection{Two-Layer Tracing Architecture}
\label{Two-Layer Tracing Architecture}
A key challenge in tracing ML I/O is that modern training frameworks use multi-process data loading. PyTorch's \texttt{DataLoader} with \texttt{num\_workers > 0} spawns worker subprocesses that read training data in parallel. Application-level tracing in the main process cannot observe these reads. Conversely, syscall-level tracing captures all I/O but lacks semantic context about which training phase caused a given operation.

We address this with a two-layer tracing architecture:

\subsubsection{Layer 1: Python IOHooks}
We monkey-patch Python's \texttt{builtins.open}, \texttt{torch.save}, \texttt{torch.load}, \texttt{os.replace}, and \texttt{os.rename} with wrapped versions that record each operation to a \texttt{TraceCollector}. Each event is tagged with the current training phase (e.g., \texttt{data\_staging}, \texttt{train\_epoch}, \texttt{checkpoint\_save}), providing semantic context that syscall tracing cannot offer. Events include wall-clock timestamps, monotonic nanosecond counters, sequence numbers, file paths, byte counts, and process/thread IDs. System and library paths are filtered to capture only application-level I/O.

\subsubsection{Layer 2: Linux strace}
We wrap the entire training process under \texttt{strace} with the \texttt{-f} (follow forks) flag, capturing every file-related syscall (\texttt{open}, \texttt{openat}, \texttt{read}, \texttt{write}, \texttt{close}, \texttt{pread64}, \texttt{pwrite64}, \texttt{lseek}, \texttt{mmap}, \texttt{fsync}, \texttt{fdatasync}, \texttt{rename}, \texttt{unlink}, \texttt{stat}, \texttt{fstat}) from all processes including DataLoader workers. Timestamps from both layers are aligned via sync markers recorded at the start and end of each run.

\subsubsection{Phase-Aware Collection}
Each training pipeline is instrumented with phase markers that delineate the stages of training: data staging, tokenization, model loading, training epochs, validation epochs, checkpointing, and model export. This enables per-phase analysis of I/O behavior and lets us distinguish, for example, the sequential writes during sharding from the bursty writes during checkpointing.

The tracer outputs one JSON file per model run containing all timestamped events, plus optional strace logs that are parsed into structured JSON for analysis.

\section{Experimental Setup}

All experiments were conducted on the NRP Nautilus Kubernetes Cluster with the following specifications:

\begin{table}[htbp]

\caption{Experimental Environment}
\begin{center}
\begin{tabular}{ll}
\toprule
\textbf{Component} & \textbf{Specification} \\
\midrule
CPU & 4 compute cores of AMD EPYC 7343 @ 3.2 GHz\\
Memory & 100 Gi \\
GPU & 1$\times$ NVIDIA RTX 3080 \\
CUDA & 13.1 \\
OS & Ubuntu 24.04.3 LTS \\
Python & 3.13 (miniconda) \\
Storage Class & rook-ceph-block (ReadWriteOnce) \\
Disk/PVC Size & 800 Gi \\
Docker Image & nvidia/cuda:13.1.1-cudnn-devel-ubuntu24.04 \\
\bottomrule
\end{tabular}
\label{tab:env}
\end{center}
\end{table}

The storage backend is Ceph \cite{weilCephScalableHighPerformancea} distributed storage system in a Kubernetes Persistent Volume Claim (PVC), which introduces network latency on every I/O operation compared to local NVMe storage. These characteristics were visible in our measurements in section \ref{Results}.

Each model was run individually with both Python-level and strace-level tracing enabled. Training hyperparameters were selected to emulate realistic workloads while keeping total runtime manageable (approximately 2 hours total across all six models).

\begin{table}[htbp]
\caption{Model Configurations}
\begin{center}
\begin{tabular}{lllr}
\toprule
\textbf{Model} & \textbf{Dataset/ RL Environment} & \textbf{Size} & \textbf{Epochs/Steps} \\
\midrule
LT & Amazon Reviews & $\sim$5K samples & 5 epochs \\
ANN & UCI Adult & $\sim$3 MB & 30 epochs \\
SNN & N-MNIST & $\sim$1 GB & 10 epochs \\
ViT & COCO 2017 & $\sim$20 GB & 10 epochs \\
Diffusion & COCO Captions & $\sim$20 GB & 10K steps \\
RL & Atari Learning Environment & N/A & 1M timesteps \\
\bottomrule
\end{tabular}
\label{tab:models}
\end{center}
\end{table}

\section{Results}
\label{Results}

\subsection{Phase-Level Time Distribution}

\begin{figure}[htbp]
\centerline{\includegraphics[width=\columnwidth]{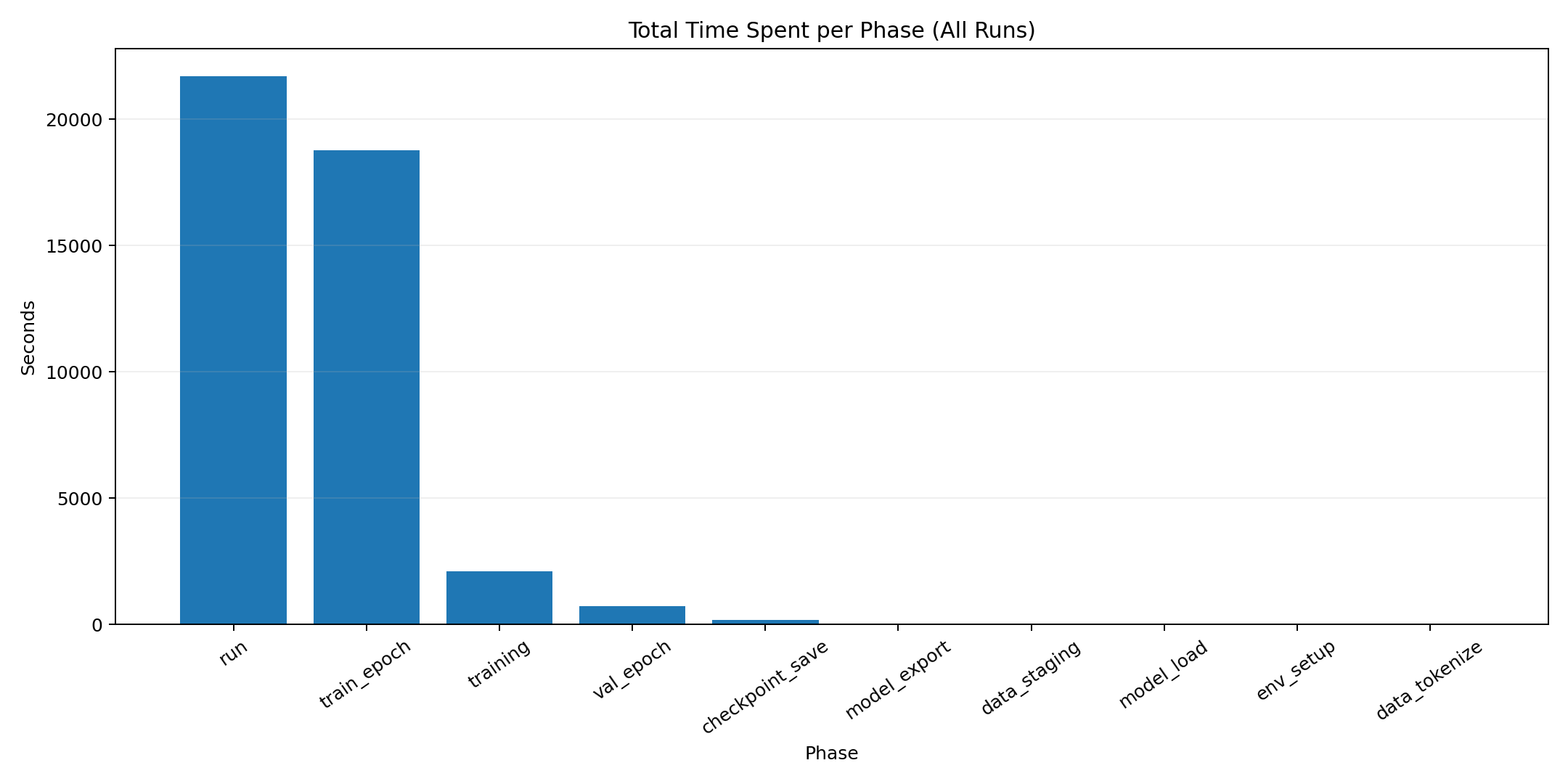}}
\caption{Total time spent per phase across all model runs. The bar "run" refers to training the RL model. Training dominates, with I/O phases occupying a small fraction of wall time.}
\label{fig:phase_time}
\end{figure}

Fig.~\ref{fig:phase_time} shows the total time spent in each phase across all model runs. Training epochs dominate wall time ($\sim$19,000 of $\sim$22,000 total seconds), confirming that ML training is fundamentally compute-bound. All I/O-intensive phases---data staging, checkpointing, model loading, tokenization, and export---collectively account for less than 15\% of total runtime.

\subsection{Python-Level vs.\ Syscall-Level Tracing}

\begin{figure}[htbp]
    \centering
    \begin{subfigure}{\columnwidth}
        \centering
        \includegraphics[width=\columnwidth]{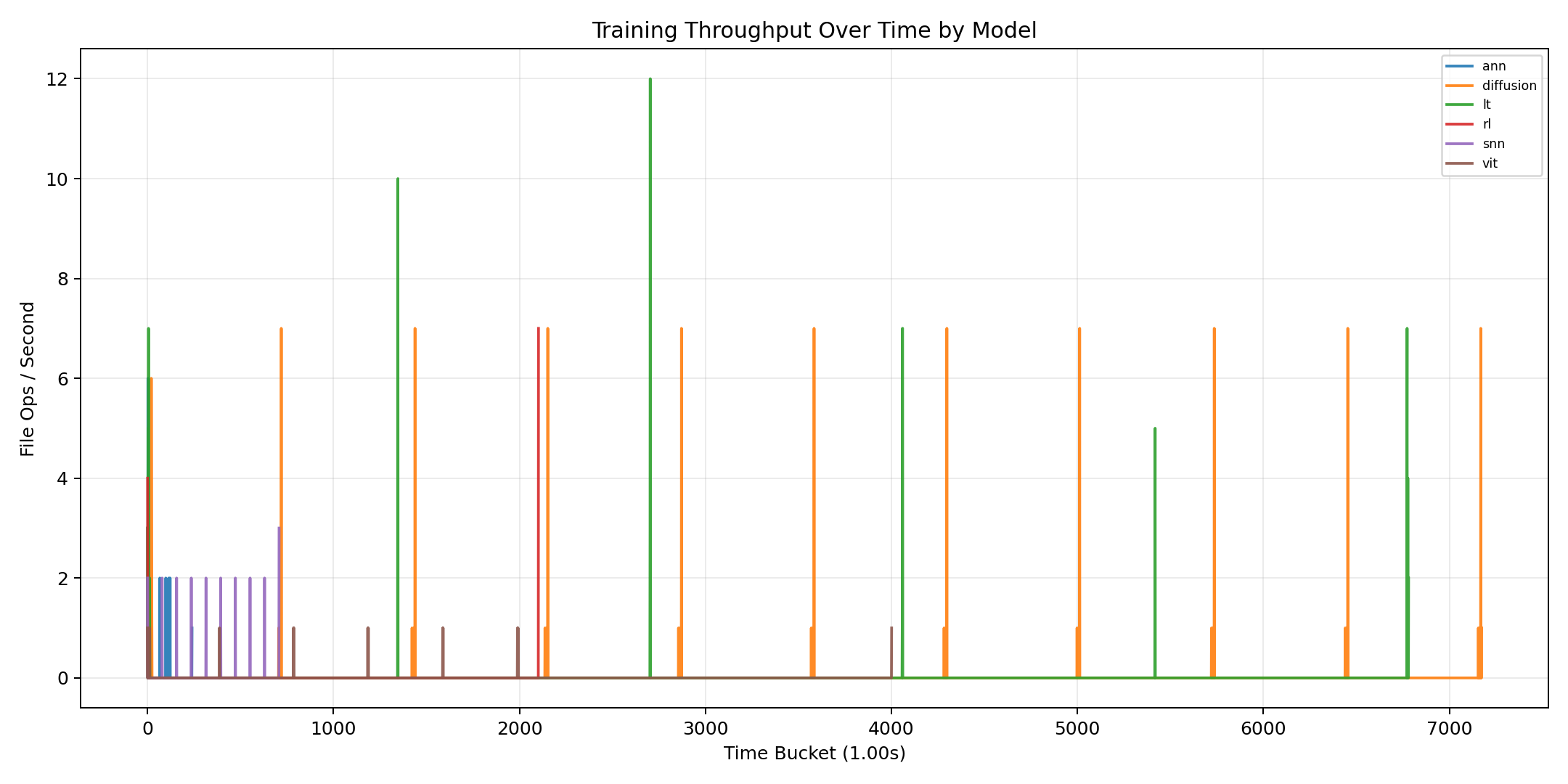}
        \caption{}
        \label{fig:training_throughput}
    \end{subfigure}
    \begin{subfigure}{\columnwidth}
        \centering
        \includegraphics[width=\columnwidth]{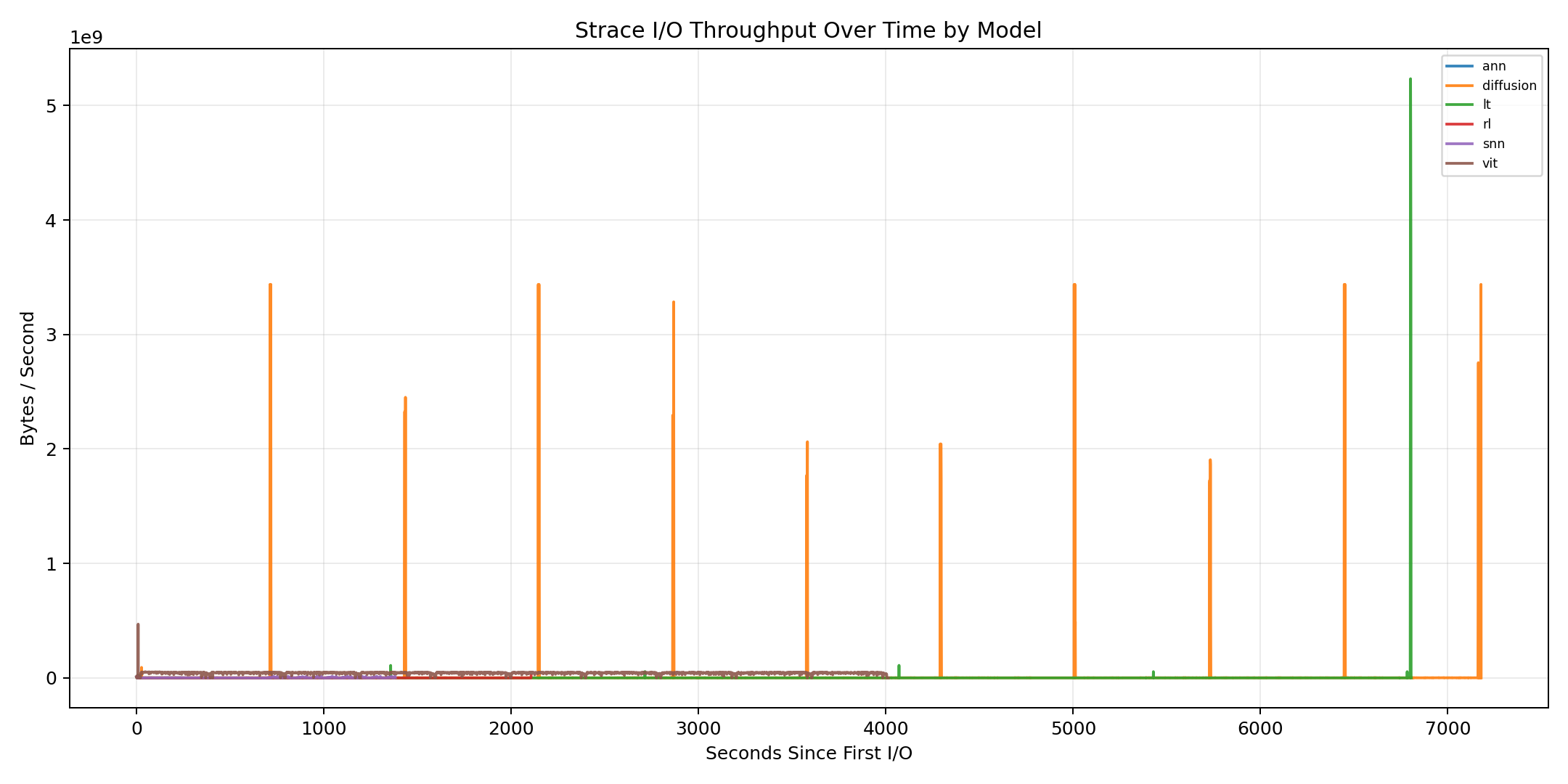}
        \caption{}
        \label{fig:strace_bytes}
    \end{subfigure}
    \caption{Python-level file ops/s (a.) vs.\ strace bytes/s (b.). Python hooks show sparse events; strace reveals multi-GB/s bursts during checkpoint writes.}
    \label{fig:python_vs_strace}
\end{figure}

Fig.~\ref{fig:python_vs_strace} compares the Python-level trace (file operations per second) with the strace-level trace (bytes per second) over the duration of the benchmark run. The Python trace shows sparse spikes at 6--12 operations per second, corresponding to checkpoint saves and model exports. Between spikes, throughput is near zero.

The strace trace reveals a dramatically different picture. Diffusion checkpoint saves produce bursts of 2--3.5~GB/s, and the LT model export generates a 5.3~GB/s spike. The Python tracer correctly identifies \emph{when} I/O occurs but dramatically understates \emph{how much} data is transferred, because each \texttt{torch.save} call appears as a single operation despite moving gigabytes of data.

\subsection{Syscall Distribution by Model}

\begin{figure}[htbp]
\centerline{\includegraphics[width=\columnwidth]{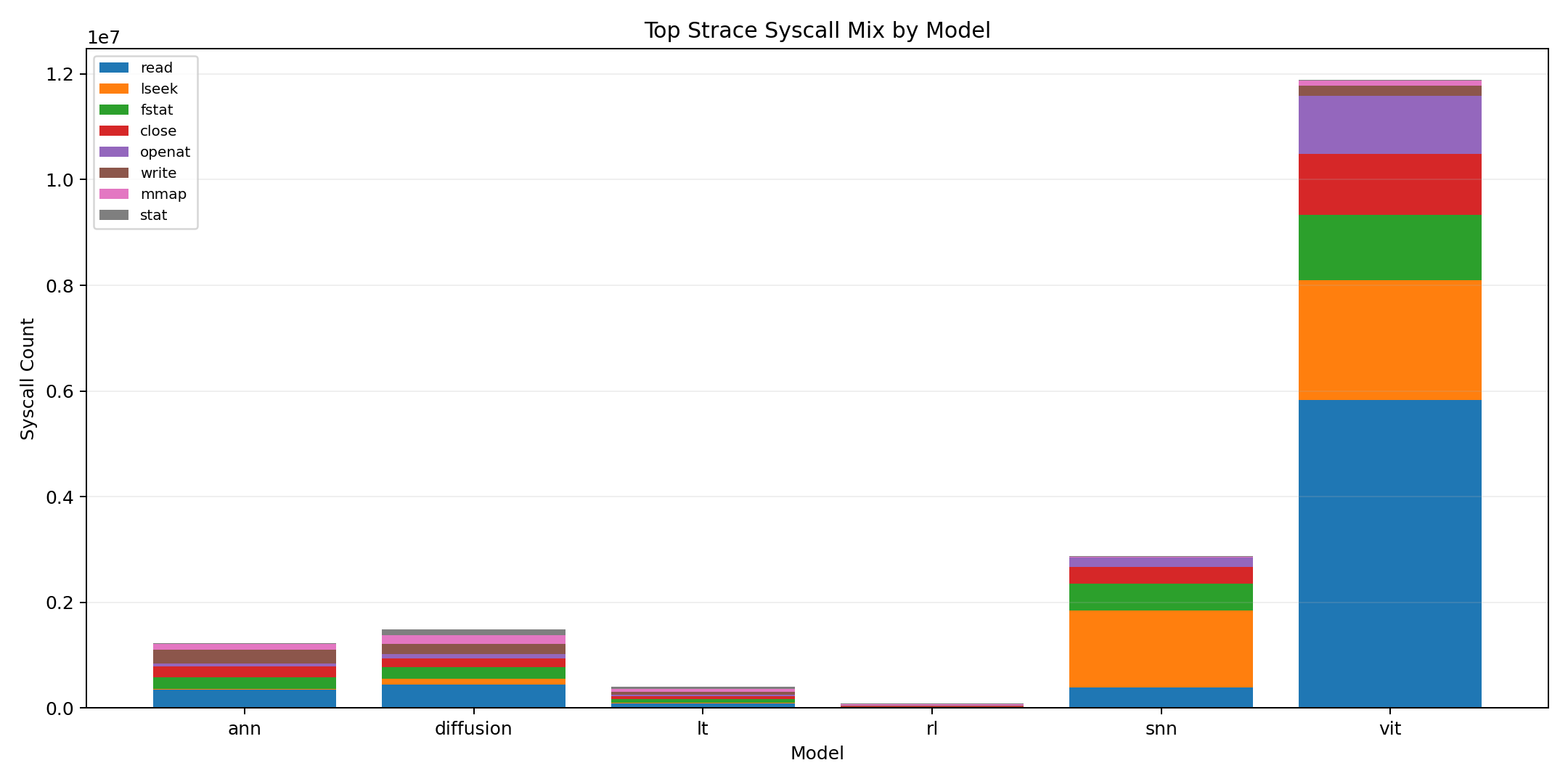}}
\caption{Syscall count by model from strace. ViT dominates with 12M syscalls from DataLoader worker image reads that are invisible to Python-level hooks.}
\label{fig:syscall_mix}
\end{figure}

Fig.~\ref{fig:syscall_mix} shows the total syscall count per model as captured by strace. ViT generates $\sim$12 million syscalls which is an order of magnitude more than any other model, and was dominated by \texttt{read} ($\sim$6M) and \texttt{lseek} ($\sim$4M) calls from DataLoader workers reading 118K JPEG images per epoch. This I/O was \emph{invisible} in the Python-level trace, which showed only $\sim$13 operations for ViT.

The path category analysis confirms that ViT's syscalls are overwhelmingly directed at the dataset cache (COCO images). SNN produces $\sim$3M syscalls from reading the N-MNIST binary files. Diffusion and ANN generate 1--1.5M syscalls each. RL produces negligible I/O, confirming that its training data is generated in memory.

The large difference in utilization between the ViT and Diffusion model is likely because the diffusion model did not use the entirety of the dataset while training within the allocated number of training steps specified, whereas the ViT trained over the dataset 10 times.

\subsection{Checkpoint I/O Dominance}

\begin{figure}[htbp]
\centerline{\includegraphics[width=\columnwidth]{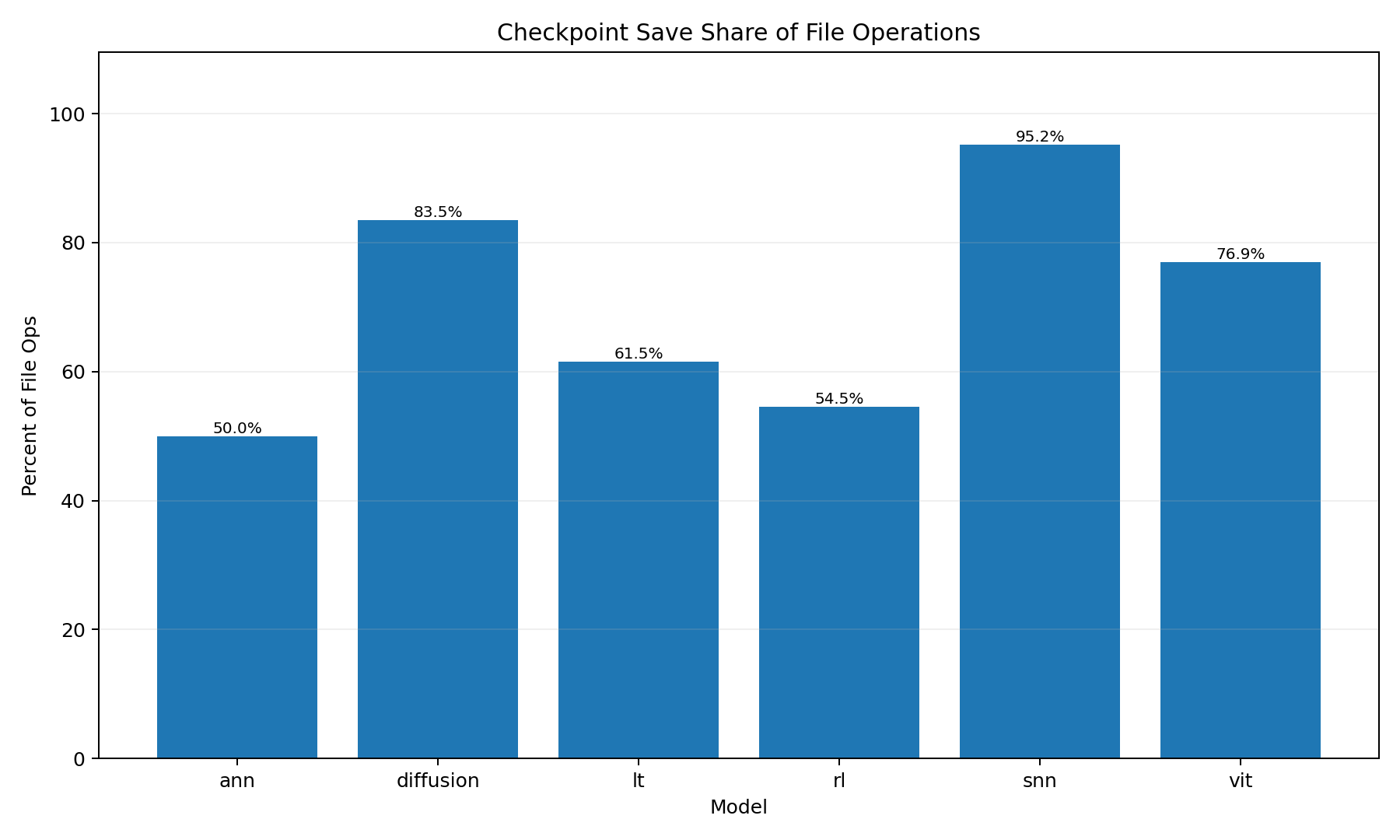}}
\caption{Checkpoint save share of total file operations by model. Checkpointing dominates Python-level I/O across all architectures.}
\label{fig:checkpoint_share}
\end{figure}

Analysis of the Python-level file operation mix reveals that checkpointing accounts for 50--95\% of all traced file operations across models (Fig.~\ref{fig:checkpoint_share}). SNN has the highest share at 95.2\%, while ANN is the lowest at 50.0\%. Diffusion writes 52 unique files during checkpointing, more than any other model, due to periodic UNet saves and Accelerator state snapshots. The \texttt{checkpoint\_save} phase alone produces 65 unique written files across all models, making it the most write-intensive phase by a wide margin.

\subsection{Syscall Latency Analysis}

\begin{figure}[htbp]
\centerline{\includegraphics[width=\columnwidth]{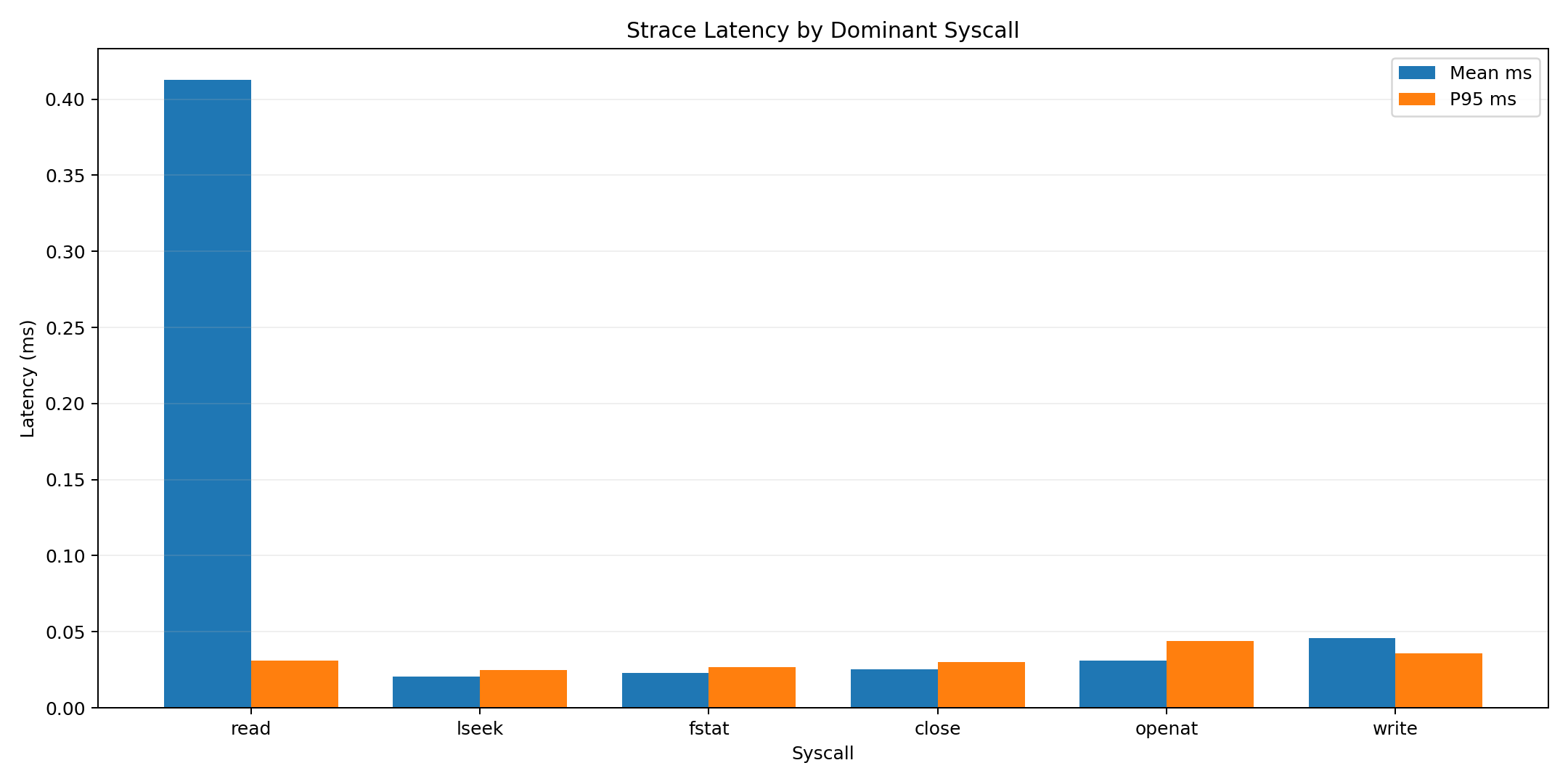}}
\caption{Mean and P95 latency by syscall type. The \texttt{read} mean (0.41~ms) 
is 13$\times$ higher than P95 (0.03~ms), indicating that rare but severe 
cache-miss reads from Ceph dominate the average despite most reads being fast.}
\label{fig:latency}
\end{figure}

Fig.~\ref{fig:latency} shows the latency distribution of syscalls from strace. The vast majority of operations complete in under 1~ms, served from the Linux page cache. However, \texttt{read} syscalls exhibit heavily skewed latency: a P95 of 0.03~ms yet a mean of 0.41~ms---13$\times$ higher---indicating that a small number of cache-miss reads hitting the Ceph storage backend are dramatically skewing the mean. These tail-latency reads occur when training data is not in the page cache and must be fetched over the network from Ceph OSDs \cite{weilCephScalableHighPerformancea}.

Other syscalls (\texttt{lseek}, \texttt{fstat}, \texttt{close}) consistently complete in $\sim$0.02~ms. \texttt{openat} shows slightly higher P95 (0.045~ms) due to occasional metadata server lookups. \texttt{write} operations average 0.045~ms, buffered by the kernel page cache, with latency spikes only during \texttt{fsync} or large checkpoint flushes.

\subsection{Byte Concentration}

\begin{figure}[htbp]
\centerline{\includegraphics[width=\columnwidth]{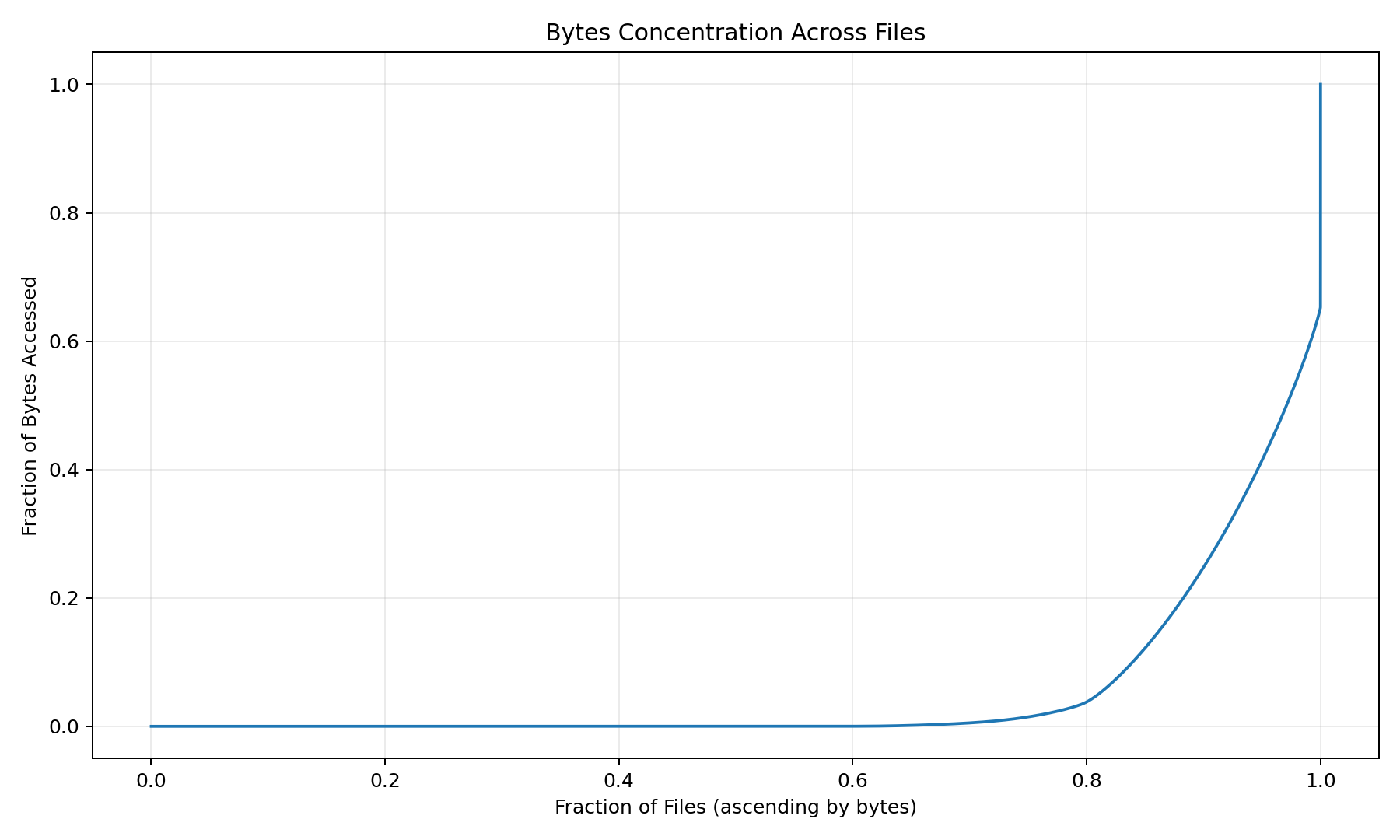}}
\caption{Bytes concentration across files (Lorenz curve). The top 10\% of files account for over 90\% of all bytes, indicating extreme concentration.}
\label{fig:concentration}
\end{figure}

Fig.~\ref{fig:concentration} shows the curve of bytes accessed across files, which matches commonly accepted data usage patterns in modern storage systems\cite{ousterhoutTraceDrivenAnalysisUNIX}. The distribution is extremely concentrated: the bottom 70\% of files by bytes accessed account for nearly 0\% of total bytes transferred. The top 10\% of files account for over 90\% of all bytes. These dominant files are checkpoint saves (860M-parameter UNet, Llama model exports), dataset archives, and Arrow shard files. The vast majority of files touched---Python libraries, configuration files, metadata---transfer negligible data.

\subsection{Temporary File Lifetimes}

The strace-level analysis reveals that ML training creates a large number of short-lived temporary files. Approximately 100,000 files have near-zero lifetime and are created, written, and renamed or deleted within milliseconds. These correspond to the atomic checkpoint pattern used by the LT pipeline (\texttt{torch.save} to a \texttt{.tmp} file followed by \texttt{os.replace}) and PyTorch's internal serialization temporaries. A file system that is slow at metadata operations (create, rename, delete) will bottleneck during checkpointing.

\section{Discussion}

\subsection{Key Findings}

Our results yield several findings relevant to storage system design for ML workloads:

\begin{enumerate}
\item \textbf{I/O is front-loaded.} Data preparation (staging, tokenization, sharding) produces the highest file operation throughput. Once data is staged, training is compute-bound with only periodic checkpoint writes. Storage systems should optimize for the burst at the beginning of training rather than sustained throughput during gradient computation.

\item \textbf{Two-layer tracing is necessary.} Python-level hooks showed ViT performing $\sim$13 file operations; strace revealed 12 million syscalls. Any I/O characterization that relies solely on application-level instrumentation will miss DataLoader worker I/O, which dominates for image-heavy models.

\item \textbf{Byte distribution follows a power law.} Fewer than 10\% of files account for over 90\% of bytes transferred. A storage system that identifies and prioritizes these hot files (checkpoints, datasets, model weights) can capture nearly all the performance benefit without optimizing for the long tail of small files.

\item \textbf{Read tail latency is the bottleneck.} Most reads complete in microseconds from page cache, but cache misses hitting Ceph add 10--100$\times$ latency. For models that re-read the same dataset every epoch (ViT, SNN), aggressive prefetching and page cache pinning would eliminate these stalls.

\item \textbf{Checkpoint writes are bursty and large.} Diffusion writes multi-gigabyte checkpoints at regular intervals. The storage system must absorb these bursts without blocking subsequent training steps. Write buffering and asynchronous flushing are critical.

\item \textbf{RL has negligible storage I/O.} Reinforcement learning generates training data at runtime in memory. This represents a fundamentally different storage profile that requires no dataset optimization.
\end{enumerate}

\subsection{Implications for Storage System Design}

Based on our findings, we propose several design considerations for ML-optimized storage systems:

\begin{itemize}
\item \textbf{Data prefetching:} Pre-stage and pin training datasets in fast storage (local NVMe or RAM cache) before training begins. The front-loaded I/O pattern means this investment pays off for the entire training duration.

\item \textbf{Checkpoint aware write path:} Implement asynchronous, buffered writes for checkpoint operations. The atomic write pattern (save to temp, then rename) should be optimized at the file system level to minimize metadata overhead. The inclusion of NVRAM and log based file systems may improve training time by allowing the checkpointing writes to occur faster, freeing up the compute to start the next training epoch.

\item \textbf{Hot-file tiering:} Automatically identify the small number of files that account for the majority of bytes transferred and place them on the fastest available storage tier.

\item \textbf{Efficient metadata operations:} The high volume of short-lived temporary files during checkpointing demands fast create/rename/delete operations. Ceph's MDS can become a bottleneck for metadata-intensive phases.
\end{itemize}

\section{Future Work}

Three directions extend this work. First, running the benchmark on different storage backends, like Lustre \cite{braam_lustre_2019}, local NVMe, HDFS \cite{shvachko_hadoop_2010}, or Amazon S3, would enable comparison of how the same workloads perform on different file systems, isolating the impact of storage architecture. Second, extending to inference workloads and training with multiple GPUs, especially GPUs distributed across multiple nodes, would characterize additional I/O profiles: inference is read-heavy with no checkpointing, while distributed training introduces parameter synchronization I/O across nodes. Third, scaling to larger models (70B+ parameter LLMs) would test whether the patterns observed here---front-loaded I/O, power-law byte concentration, bursty checkpoints---hold at production scale. Lastly, improvements can be made to NIO Bench to collect GPU information such as GPU utilization through time, which might provide insight into whether the GPU is data starved during training.

\section{Conclusion}

We presented a benchmarking framework, Neural I/O Benchmark (NIO Bench), for characterizing storage system I/O during ML training across six model architectures. Our two-layer tracing approach which combined Python-level hooks for semantic context with strace for complete syscall coverage, revealed that application-level tracing alone misses the majority of I/O for models using multi-process data loading. Testing on a Kubernetes cluster with Ceph file system, we found that I/O concentrates in data preparation phases, training is compute-bound, byte distribution follows an extreme power law, and read tail latency from distributed storage is the primary storage bottleneck. These findings provide concrete guidance for designing storage systems optimized for modern ML workloads.

\section*{Acknowledgment}

We thank the NRP Nautilus cluster for providing compute resources and the UC Santa Cruz Baskin School of Engineering for supporting this research.

\bibliographystyle{IEEEtran}
\bibliography{bib}

\end{document}